\documentclass[12pt]{article}
\usepackage{scicite}

\usepackage[utf8]{inputenc} %unicode support
\usepackage{graphicx}
\usepackage{ulem}
\usepackage{listings}
\usepackage{url}

\def\apjs{\ {ApJS}\ }

\def\apgt{\ {\raise-.5ex\hbox{$\buildrel>\over\sim$}}\ }
\def\aplt{\ {\raise-.5ex\hbox{$\buildrel<\over\sim$}}\ }
\def\lteq{\ {\raise-.5ex\hbox{$\buildrel<\over-$}}\ }

\title{Computational astrophysics for the future: \\
  {\small An open,
    modular approach with agreed standards would facilitate
    astrophysical discovery}}

\author
{Simon Portegies Zwart$^{1,\ast}$
\\
\normalsize{$^{1}$Leiden Observatory, Leiden University, PO Box 9513, 2300 RA, Leiden, The Netherlands}\\
\\
\normalsize{$^\ast$E-mail:  spz@strw.leidenuniv.nl}
}
\date{}
\begin{document} 
\maketitle 

Scientific discovery is mediated by ideas that, after being formulated
in hypotheses, can be tested, validated, and quantified before they
eventually lead to accepted concepts. Computer-mediated discovery in
astrophysics is no exception, but antiquated code that is only
intelligible to scientists who were involved in writing it is holding
up scientific discovery in the field. A bold initiative is needed to
modernize astrophysics code and make it transparent and useful beyond
a small group of scientists.  Scientific software is like a prototype
in a laboratory experiment; it must stimulate experimentation. The
eventual code is a description of concepts and their relationships,
which are imperative for reproducibility and validating the
results. In recent decades, hardware performance and computer
languages have both changed dramatically. Software engineering has
also experienced major advances, such as multi-lingual implementations
and the introduction of design patterns. Industry and large scientific
endeavors, such as CERN, have embraced these changes; in addition,
their software is built on various assisting packages to vitalize
modularity, platform independency, and message brokering. Although
leading to a better structure, these developments also lead to higher
complexity \cite{295895} and a sharp increase in the number of code
lines \cite{Lehman:1997:MLS:823454.823901}. The latter is
particularly noticeable in the growth of the Linux kernel \cite{2015Spinellis}.
Astronomical source code remains tiny by industrial standards, and its
structure is characterized by developments during the “software
crisis” of 1965 to 1985, when software was written as a long list of
instructions without formal structure \cite{NATO1969}. The relative simplicity of
these “dinosource” codes facilitates their survival but frustrates
further development. Scientific source code is experimental in much
the same way as laboratory experiments; it is not a final concept and
is never ideally organized because it is intended to mediate
exploration. By contrast, industrial code is mature but restrains
experimentation. Furthermore, industry can afford dedicated teams to
design and maintain software, whereas astronomical software
development is organized in indigent “families” of researchers. A
modular approach with agreed standards is essential to embolden
astrophysical discovery by computer.

\section{CITING AND SHARING}

The simple structure of astronomical software packages has enabled
them to survive, some even since the 1970s. Although this dinosource
code is often written in an ancient dialect, the underlying physics
has hardly changed. About 58\% of these codes are publicly available
\cite{2018ApJS..236...10A}, but many are obfuscated because they have
received multiple updates in the form of patches, keeping the
deprecated code for backward compatibility. Existing code therefore
forms no suitable basis for building a general framework.  We know
little about undisclosed codes, but it is unlikely that they are
different.  Authors may decide not to distribute their source code;
this is generally motivated by a desire to preserve a head start on
the competition, or for fear that erroneous results produced by others
discredit their efforts. Such secretly developed codes are of no help
to the community and produce unverifiable results. In experimental
sciences, failure to publish laboratory details is equivalent to
failure in properly describing an experiment
\cite{Collberg:2016:RCS:2897191.2812803}, and this cannot be much
different for source code. But so long as public codes are copied,
adapted, or expanded and eventually released under a different name,
one can hardly expect the community to change its behavior. The
apparent lack of trust that sufficient credit is given to code
development can only be lifted by enabling codes to be cited directly.
The Astrophysics Source Code Library \cite{2013ASPC..475..387A}
provides a platform that mediates sharing and citing of astronomical
source code, but it does not assign a digital object identifier
(DOI). Such a service is provided by Zenodo\footnote{see Zenodo:
  \url{www.zenodo.org}.}, and all ingredients to organize the
community are in principle available.

A specialized journal that assigns a DOI to a code, publishes
simulation data, and enables discussion could be the solution. Such
services are offered by the journal Geoscientific Model Development
\footnote{see Geoscientific Model Development:
  \url{www.geoscientific-model-development.net}.}, but not for the
astronomical community.

\section{TRANSPARENCY AND CONSISTENCY}

Although citing and sharing are important, the real breakthrough will
come with the introduction of a common interface and data-exchange
protocol. Very few astronomical simulation codes read each other's
data formats or parse each other’s input parameters.  This lack of
consistency hinders information exchange from one implementation to
another and impedes its use by non-family researchers. The introduction
of a standard for data exchange will eventually lead to more stable
software, fewer bugs, and improved validation, verification, and
quantification of the results. If only the community could agree on
such a standard. Courageous attempts to achieve these objectives
include the Virtual Observatory \footnote{see
  \url{http://www.ivoa.net/documents/Notes/IVOAArchitecture}, in: IVOA
  Note 23/11, 2010.} for the observational community, Astropy
\cite{collaboration2013astropy} for data analysis, and the
Astrophysical Multipurpose Software Environment (AMUSE) \cite{AMUSE}
for multiscale simulations. However, the first two of these are not
modular, and the third is based on existing community code.

\section{MODULARITY AND TRANSPOSABILITY}
\begin{figure}
\includegraphics[width=1.0\columnwidth]{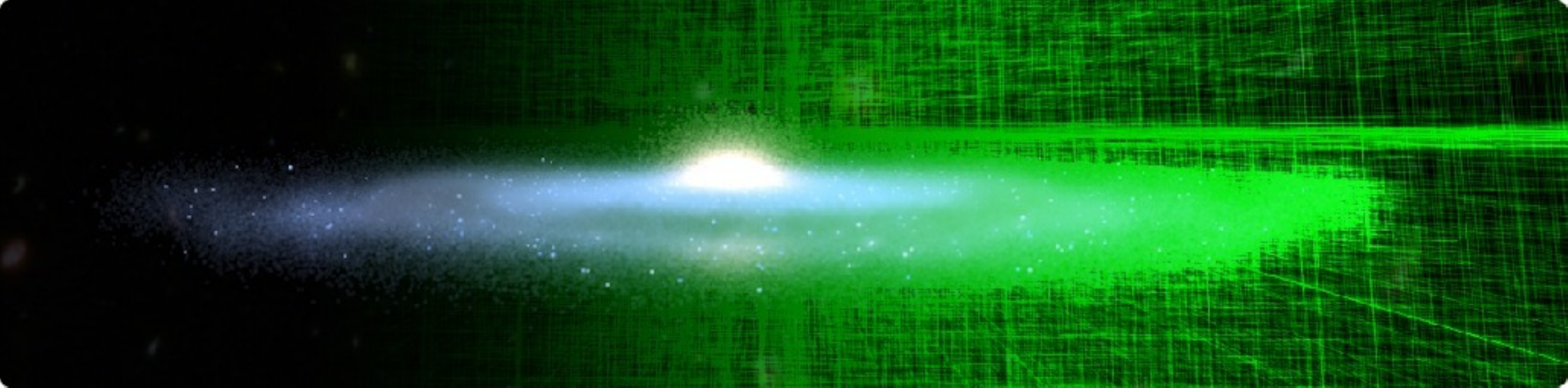}
\caption{Results from a simulation of the Milky Way Galaxy using the
  Bonsai code on the ORNL Titan supercomputer
  \cite{Bedorf:2014:PGT:2683593.2683600}. The left-hand side has been
  modified to look more like the observations; the right-hand side
  shows the intrinsic data structure at runtime. Bonsai is a highly
  optimized but small and dedicated code for a specific application.
  \label{figure}
}
\end{figure}

One potential analog for a modular astrophysical software system is
provided by LEGO bricks, which have a general interface and abstract
fine-grained structure that enable endless permutations. For
scientific software, however, it is insufficient to design a common
interface and establish a modular structure, because the current
implementations are too complex to be used in an aggregate
environment. Rather than expanding existing code, one can envision the
reimplementation of libraries of fundamental solvers that are designed
to cooperate. The resulting shorter code has the advantage of being
easier to understand, and it contains fewer bugs
\cite{McConnell:2004:CCS:1096143}. It may seem overkill to rewrite
code from scratch, but industry is in a continuous cycle of
re-engineering.  For general programming, however, the fine-graining in
the analogy with LEGO is bound to introduce a variety of optimization
problems \cite{Brooks:1995:MM:207583}. To make the LEGO concept for
software design work, the subdivision in minimal routines should not
be too rigorous. Instead of dividing the software into its fundamental
operations, a more coarse grained structure could allow collections of
fundamental elements to form parts of a larger structure. Twenty years
after its introduction, LEGO also realized that many of its young
users desire less fine-grained design with more recognizable
shapes. The resulting DUPLO blocks reduced graininess while mediating
and stimulating experimentation. In terms of software, they would be
equivalent to self-contained objects with a common interface
structure, which enables inter-object operations.  From a numerical
point of view, such an environment should include code-exchange and
data-exchange strategies as well as method-coupling paradigms. With
the DUPLO analogy, building blocks become the essential ingredient for
individual physical domain–specific solvers. These can be easily
tested and validated using a daily executed standard test through
online services such as Travis CI\footnote{see
  \url{https://travis-ci.org}.}. Combinations or
communication-essential parts that frequently appear could be
reimplemented as separate modules to improve performance and
scalability (see the figure\,\ref{figure}). The value per module
increases as the environment grows, which again stimulates other
researchers to contribute.  Independently working dedicated snippets
should be publishable and citable—for example, in a code-dedicated
journal

\section{Outlook}

We have reached the point where it is no longer reasonable to expect a
single graduate student to develop a production-quality astronomical
software package. This results in the loss of individual contributions
and the further expansion of existing dinosource packages. With a
modular approach, a general framework can be complemented by
reimplementing dedicated codes that solve only part of the
astrophysical problem, each of which can be credited separately. A
standardized environment would also stimulate collaboration, mediate
validation and verification, expand the scientific life span of
individual contributions, and allow the development and maintenance
effort to be absorbed by the community. Without the community taking
responsibility in establishing a more collaborative environment, the
lack of evolution in the engineering of astronomical software will
lead to its devolution.

\centering

\section*{Acknowledgements}
I thank J. B\'edorf, A. Hoekstra, and A. van Elteren for discussions.

\end{document}